\documentclass[conference]{IEEEtran}
\usepackage{booktabs} % For formal tables
\usepackage{algorithm}
\usepackage{algpseudocode}
\usepackage{color,colortbl}
\usepackage{array}
\usepackage{amsfonts}
\usepackage{tikz}
\usepackage{bm}
\usepackage{pgfplots}
\pgfplotsset{compat=1.14}
\usepackage{mathtools, cuted}
\ifCLASSOPTIONcompsoc \usepackage[caption=false,font=normalsize,labelfont=sf,textfont=sf]{subfig}
\else
\usepackage[caption=false,font=footnotesize]{subfig}
\fi
\usepackage{calligra}
\usepackage{url}
\usepackage{multirow}
\usepackage{float}
\usepackage{paralist}
\usepackage{tikz,pgfplots,pgfplotstable}
\newcommand{\tacc}{{\tt TACC\_Stats}}
\newtheorem{defn}{Definition}[section]

\begin{document}
\title{Tracking System Behavior from Resource Usage Data}
\author{\IEEEauthorblockN{Niyazi Sorkunlu}
\IEEEauthorblockA{Computer Science and Engineering \\
  University at Buffalo, \\
  State University of New York\\
  Buffalo, New York 14260\\
Email: niyaziso@buffalo.edu}
\and
\IEEEauthorblockN{Varun Chandola}
\IEEEauthorblockA{Computer Science and Engineering \\
  University at Buffalo, \\
  State University of New York\\
  Buffalo, New York 14260\\
Email: chandola@buffalo.edu}
\and
\IEEEauthorblockN{Abani Patra}
\IEEEauthorblockA{Mechanical and Aerospace Engineering \\
  University at Buffalo, \\
  State University of New York\\
  Buffalo, New York 14260\\
Email: abani@buffalo.edu}}
\maketitle
\begin{abstract}
  Resource usage data, collected using tools such as TACC\_Stats, capture the resource utilization by nodes within a high performance computing system. We present methods to analyze the resource usage data to understand the system performance and identify performance anomalies. 
  The core idea is to model the data as a three-way tensor corresponding to the compute nodes, usage metrics, and time. Using the reconstruction error between the original tensor and the tensor reconstructed from a low rank tensor decomposition, as a scalar performance metric, enables us to monitor the performance of the system in an online fashion. This error statistic is then used for anomaly detection that relies on the assumption that the normal/routine behavior of the system can be captured using a low rank approximation of the original tensor. We evaluate the performance of the algorithm using information gathered from system logs and show that the performance anomalies identified by the proposed method correlates with critical errors reported in the system logs. Results are shown for data collected for 2013 from the Lonestar4 system at the Texas Advanced Computing Center (TACC).
\end{abstract}
\IEEEpeerreviewmaketitle
\section{Introduction}
\label{sec:introduction}
Modern high performance computing (HPC) systems play a critical role in advancing scientific and engineering research. Given the high costs associated with the operation of such systems, it is vital that they are utilized as efficiently as possible. System level monitoring plays an important role in understanding the performance of such systems to ensure smooth and efficient operations.

%An estimated 20\% or more of computing capacity in a large HPC facility is wasted due to software failures and errors~\cite{Cappello2014}. This issue is even worse if we consider the waste due to poorly written applications running on such systems. In the future exascale computing world, this situation is expected to become even more exacerbated. 
Several tools exist for collecting and visualizing resource usage data from large scale HPC installations (e.g. Texas Advanced Computing Center \tacc~\cite{Evans2014}, XSEDE Metrics on Demand or XDMoD~\cite{Palmer2015}, etc.). Such tools can produce large amounts of high dimensional resource usage data at a high temporal frequency for each computational node. The data collected by such tools provides a real-time view of the performance of the system and can be analyzed to identify system performance anomalies.

Existing methods typically monitor each system resource or state for every compute node independently for potential deviations or anomalies~\cite{Peiris:2014}, relying on a pre-defined threshold. Such methods can identify only those anomalous scenarios in which an individual node exhibits significant deviation for individual resources. These methods often miss anomalies that are under the threshold. Such anomalies, are weakly manifested across several nodes and multiple system resources and can be potentially detected by understanding the interactions between the different aspects of the system.

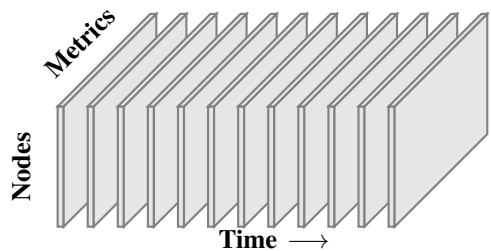
\begin{figure}	
\centering
\begin{tikzpicture}[thick, scale=0.8]
  \pgfmathsetmacro{\cubex}{0.1}
  \pgfmathsetmacro{\cubey}{2}
  \pgfmathsetmacro{\cubez}{4}
  \foreach \i in {1,...,12}
  {
  \pgfmathsetmacro{\posx}{-6 + 0.5*\i}
  \draw[gray,fill=gray!20] (\posx,0,0) -- ++(-\cubex,0,0) -- ++(0,-\cubey,0) -- ++(\cubex,0,0) -- cycle;
  \draw[gray,fill=gray!20] (\posx,0,0) -- ++(0,0,-\cubez) -- ++(0,-\cubey,0) -- ++(0,0,\cubez) -- cycle;
  \draw[gray,fill=gray!20] (\posx,0,0) -- ++(-\cubex,0,0) -- ++(0,0,-\cubez) -- ++(\cubex,0,0) -- cycle;
  }

  \draw(-6.2,-1,0) node[rotate=90]{\bf Nodes};
  \draw(-6,0.3,-2.0) node[rotate=45]{\bf Metrics};
  \draw(-2,-2.2,0) node{\bf Time $\longrightarrow$};
\end{tikzpicture}
\caption{Conceptual representation of resource usage data as a 3-way tensor consisting of compute {\bf nodes}, usage {\bf metrics} and {\bf time} dimensions as lateral slices. Each dimension is referred to as a {\em mode} of the tensor.}
\label{fig:performancecountertensor}
\end{figure}

We present a method that operates on resource usage data, such as hardware performance counters, filesystem operational counts, network device usage, etc., to produce a single error metric, that can be used to track the peformance of the HPC system over time. In particular, the method models the multi-dimensional resource usage metrics available for the entire system, as a three way tensor (a generalization of a two-way matrix), as shown in Figure~\ref{fig:performancecountertensor}. The three dimensions of the tensor correspond to the compute nodes, performance metrics associated with system resources (cpu load, memory usage, etc.), and time. We argue that the behavior of an HPC system will exhibit correlations across the different dimensions, at least within a short temporal neighborhood. Hence, the tensor representing the resource usage data can be reasonably represented using a lower rank tensor approximation. 
\begin{figure*}[htbp]
  \centering
  \includegraphics[width=0.9\textwidth]{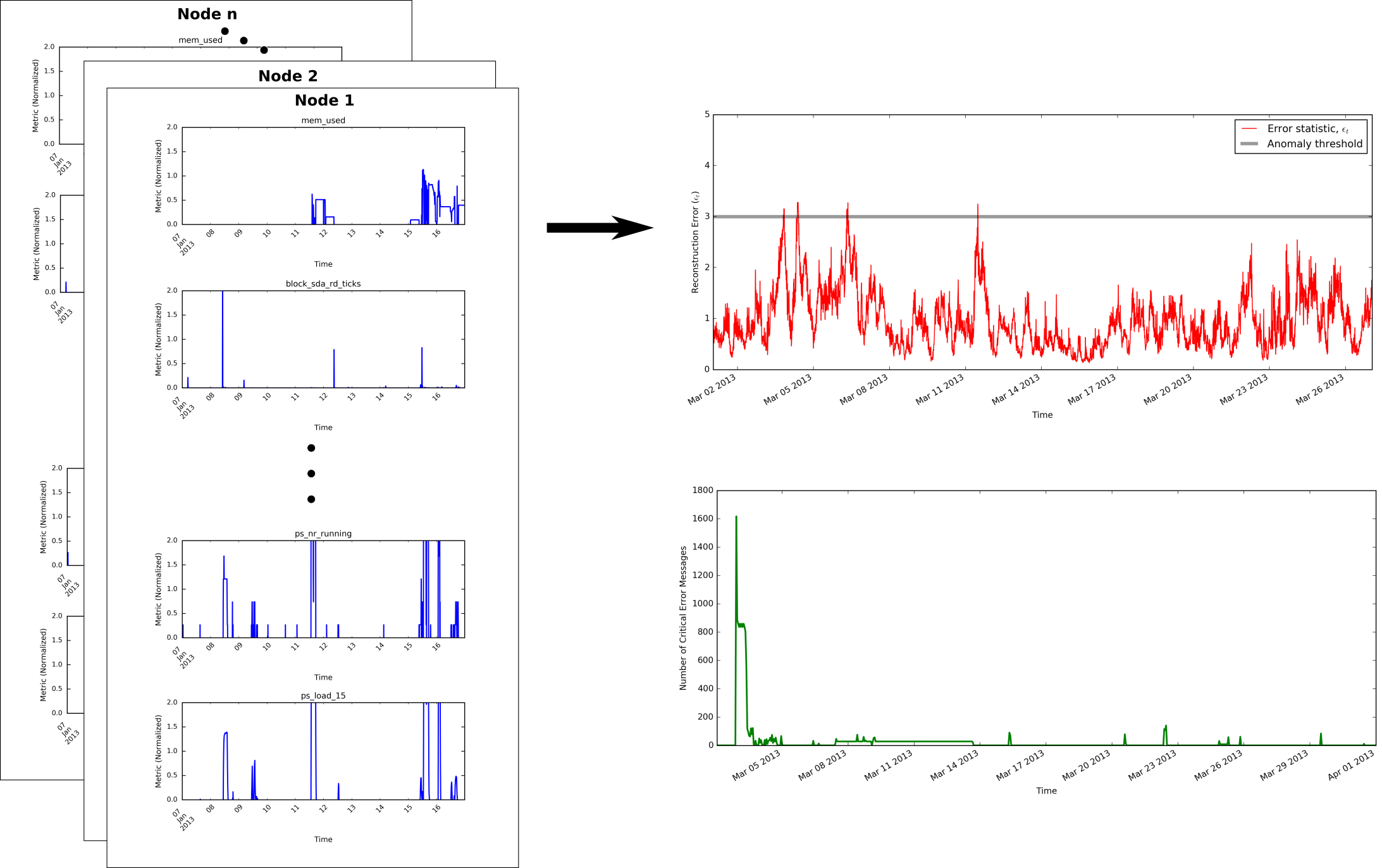}
  \caption{Workflow for the proposed methodology. Resource usage data from multiple compute nodes (first panel) is combined as a tensor to produce an error metric (second panel). The anomalous spikes are shown to correspond to reported errors extracted from message logs. An example of such messages is shown in green (third panel).}
  \label{fig:overview}
\end{figure*}
Similar arguments have been posed in the area of network anomaly detection, where the data is typically modeled as a 2-way matrix, instead of a tensor~\cite{Huang:2006}. We empirically demonstrate the correlations in the resource usage data and show how a significantly lower rank approximation of the tensor can capture most of the information in the original tensor. Moreover, we observe that the residual information, or the {\em residual error}, is reasonably stable under the assumption of normal system behavior. 

We base our algorithm on the key assumption that when the system is exhibiting anomalous behavior, in terms of performance, the residual error can be used as an indicator for the anomaly. The overall methodology for identifying performance anomalies is illustrated in Figure~\ref{fig:overview}. The method combines the multi-dimensional, multi-node, resource usage data into a univariate error time series. This error time series can be used as an anomaly signal to identify performance anomalies. Experiments results are shown using data obtained for the Lonestar4 cluster at the Texas Advanced Computing Center (TACC). Error information from application level logs is used to validate the anomalies identified by the proposed solutions. %Further study reveals an interesting relationship between the anomalies identified from the resource usage data and the application level logs. In particular, we notice a {\em lagged} relationship between the two.

\section{System Model}
\label{sec:data}
The system under consideration in this paper is a typical high performance cluster consisting of a set of compute nodes running jobs or tasks at any time instance on one or more nodes. The {\em resource usage} data, used in this paper, is collected by monitoring tools such as the \tacc. The monitoring tool collects usage statistics and hardware performance counters, collectively referred to as {\em metrics} for each node at a given time instance. See Table~\ref{tab:metrics} for a list of usage metrics used in our experiments. This data is collected either at the beginning or end of a job, or periodically via a scheduled collection. This data can be represented as a $N \times M$ matrix, denoted as ${\bf T}_t$, where $N$ is the number of nodes, $M$ is the number of metrics, and the subscript $t$ is the time at which the data was recorded. Thus, ${\bf T}_t[n][m]$ is the value of the $m^{th}$ metric for the $n^{th}$ node\footnote{For metrics that correspond to running counters, we consider the difference between values for successive time instances.}. At the same time, the data also provides job specific information, i.e., what jobs are running at each node at a given time instance $t$. The job information is aggregated for a time window, instead of a specific time instance, as explained below.
\subsection{Tensor Representation of Resource Usage Data}
In our analysis, we consider data for a time window instead of a specific time instance. Considering only the periodically collected data (typically collected in intervals of 10 minutes), we define a 3 way tensor, denoted as $\mathcal{T}_t \in \mathbb{R}^{N\times T \times M}$, where $T$ is the number of time samples available in a fixed sized time window. The subscript $t$ denotes the starting time instance for this time window. Each slice in the tensor $\mathcal{T}_t$ is essentially the usage matrix defined earlier, ${\bf T}_{t'}, t \le t' \le t+T$. This is illustrated in Figure~\ref{fig:performancecountertensor}.
\subsection{Job Matrix for a Time Window}
While the tensor for a given time window only considers the periodically collected usage data, we use the data collected at start and end of every job, to construct a job matrix ${\bf J}_t \in \mathbb{R}^{N \times J_t}$ for a time window starting at time $t$. $J_t$ is the total number of active jobs in the system for that time window. Each entry, ${\bf J}_t[n,j]$, is set to 1 if the $n^{th}$ node was running the $j^{th}$ job during that time window, otherwise it is set to 0.  

\section{Tensor Decomposition Based Error Statistic}
\label{sec:error}
One reason for using a tensor based representation of the resource usage data is to be able to exploit the correlations among the different modes in the data. For instance, several usage metrics (See Table~\ref{tab:metrics}) are expected to exhibit a correlated structure. Similarly, one would expect a temporal regularity in the data and correlations across different computational nodes. 
We exploit the presence of these structures by employing tensor decomposition methods to express the data in terms of low rank factorization of the previously defined tensor representation for a given time window. 
Two types of error statistics are defined using the low rank approximation of the tensor. We first provide a short background on tensor decomposition, which will then be used to obtain the statistics.
\subsection{Background - Tensor Decomposition}
Two of the most widely used tensor decomposition methods are: the CANDECOMP/PARAFAC or CP method~\cite{CEM:CEM582} and Tucker decomposition~\cite{Tucker1966}. 
%In this paper, we use the CP method for tensor decomposition, which works as follows:
Both methods are higher order generalizations of the widely used Singular Value Decomposition (SVD) and Principal Component Analysis (PCA), respectively. We provide brief description of the two methods here.
\begin{figure}	
  \centering
  \subfloat{Tucker}{
    \begin{tikzpicture}
  \draw[white,fill=gray!40] (0,0,0) -- ++(-2,0,0) -- ++(0,-1,0) -- ++(2,0,0) -- cycle;
  \draw[white,fill=gray!40] (0,0,0) -- ++(-2,0,0) -- ++(0,0,-1) -- ++(2,0,0) -- cycle;
  \draw[white,fill=gray!40] (0,0,0) -- ++(0,0,-1) -- ++(0,-1,0) -- ++(0,0,1) -- cycle;
  \node at (-1,-0.5) {$\mathcal{T}$};
  \node at (2/3,-1/3) {$\approx$};
  \draw[white,fill=gray!40] (1.5,0.25,0) -- ++(-0.5,0,0) -- ++(0,-1,0) -- ++(0.5,0,0) -- cycle;
  \node at (1.25,-0.25) {\footnotesize ${\bf A}^M$};
  \draw[white,fill=gray!40] (3,0,0) -- ++(-1,0,0) -- ++(0,-0.5,0) -- ++(1,0,0) -- cycle;
  \draw[white,fill=gray!40] (3,0,0) -- ++(-1,0,0) -- ++(0,0,-0.5) -- ++(1,0,0) -- cycle;
  \draw[white,fill=gray!40] (3,0,0) -- ++(0,0,-0.5) -- ++(0,-0.5,0) -- ++(0,0,0.5) -- cycle;
  \node at (2.5,-0.25) {$\mathcal{G}$};
  \draw[white,fill=gray!40] (3.5,.5,0) -- ++(-2,0,0) -- ++(0,0,-1) -- ++(2,0,0) -- cycle;
  \draw[white,fill=gray!40] (3.5,0,0) -- ++(0,0,-1.0) -- ++(0,-0.5,0) -- ++(0,0,1.0) -- cycle;
  \node at (2.75,0.75) {\footnotesize ${\bf A}^T$};
  \node at (3.7,-0.15) {\tiny ${\bf A}^N$};
\end{tikzpicture}
    \label{fig:tucker}		
  } 
  \subfloat{CP}{
    \begin{tikzpicture}
  \draw[white,fill=gray!40] (0,0,0) -- ++(-2,0,0) -- ++(0,-1,0) -- ++(2,0,0) -- cycle;
  \draw[white,fill=gray!40] (0,0,0) -- ++(-2,0,0) -- ++(0,0,-1) -- ++(2,0,0) -- cycle;
  \draw[white,fill=gray!40] (0,0,0) -- ++(0,0,-1) -- ++(0,-1,0) -- ++(0,0,1) -- cycle;
  \node at (-1,-0.5) {$\mathcal{T}$};
  \node at (2/3,-1/3) {$\approx$};
  \node at (1,-1/3) {$\lambda_1$};

  \draw[white,fill=gray!40] (1.4,0,0) -- ++(-0.2,0,0) -- ++(0,-1,0) -- ++(0.2,0,0) -- cycle;
  \draw[white,fill=gray!40] (3.5,0,0) -- ++(-2,0,0) -- ++(0,-0.2,0) -- ++(2,0,0) -- cycle;
  \draw[white,fill=gray!40] (1.4,0,-0.1) -- ++(-0.2,0,0) -- ++(0,0,-1) -- ++(0.2,0,0) -- cycle;
  \node at (2.5,-1.2) {$+$};
  \node at (1,-1/3-1.8) {$\lambda_2$};

  \draw[white,fill=gray!40] (1.4,-1.8,0) -- ++(-0.2,0,0) -- ++(0,-1,0) -- ++(0.2,0,0) -- cycle;
  \draw[white,fill=gray!40] (3.5,-1.8,0) -- ++(-2,0,0) -- ++(0,-0.2,0) -- ++(2,0,0) -- cycle;
  \draw[white,fill=gray!40] (1.4,-1.8,-0.1) -- ++(-0.2,0,0) -- ++(0,0,-1) -- ++(0.2,0,0) -- cycle;
  \node at (2.5,-2.7) {$+$};
  \node at (2.5,-3.0) {$\vdots$};
  \node at (2.5,-3.5) {$+$};
  \node at (1,-1/3-4) {$\lambda_n$};

  \draw[white,fill=gray!40] (1.4,-3.8,0) -- ++(-0.2,0,0) -- ++(0,-1,0) -- ++(0.2,0,0) -- cycle;
  \draw[white,fill=gray!40] (3.5,-3.8,0) -- ++(-2,0,0) -- ++(0,-0.2,0) -- ++(2,0,0) -- cycle;
  \draw[white,fill=gray!40] (1.4,-3.8,-0.1) -- ++(-0.2,0,0) -- ++(0,0,-1) -- ++(0.2,0,0) -- cycle;
\end{tikzpicture}
    \label{fig:cp}
  }
  \caption{Tensor decomposition of a 3-way tensor using Tucker and CP}\label{fig:tensordecomp}
\end{figure}
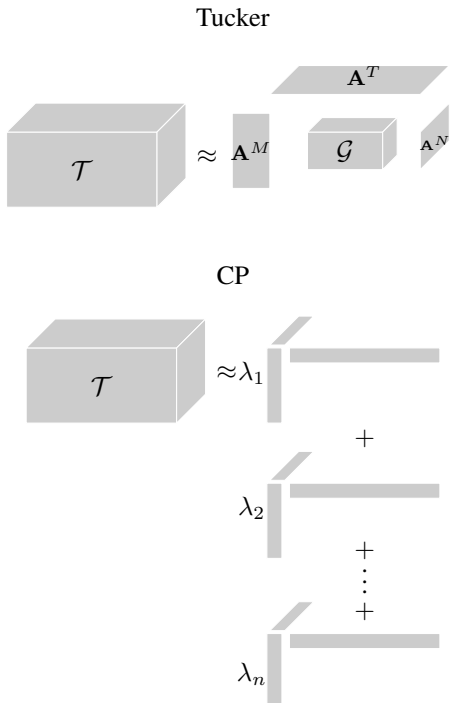

\subsubsection{Tucker Tensor Decomposition}
The Tucker decomposition~\cite{Tucker1966} is a higher order generalization of PCA and involves decomposition of a tensor into a lower rank core tensor, multiplied by a matrix along each mode as illustrated in Figure~\ref{fig:tucker}.
\begin{defn}
  Let $\mathcal{T}$ be a tensor of size $N\times T \times M$. A rank-($N',T',M'$) Tucker decomposition of $\mathcal{T}$ yields a core tensor $\mathcal{G}$ of size $N'\times T' \times M'$ and three factor matrices (usually orthogonal), ${\bf A}^N$, ${\bf A}^T$, and ${\bf A}^M$, of sizes $N \times N'$, $T \times T'$, and $M \times M'$, respectively, such that:
  \begin{equation}
  \mathcal{T} \approx \mathcal{G} \times {\bf A}^N \times {\bf A}^T \times {\bf A}^M
    \label{eqn:tucker}
  \end{equation}
  \label{defn:tucker}
\end{defn}
$(N',T',M')$ is referred to as the {\em rank} of the tensor.
The core tensor and the factor matrices capture the level of interaction between different modes. The Tucker decomposition is typically computed by solving the following optimization problem:
\begin{equation}
    \underset{\mathcal{G},{\bf A}^N,{\bf A}^T,{\bf A}^M}{\min} \Vert\mathcal{T} - \mathcal{G} \times {\bf A}^N \times {\bf A}^T \times {\bf A}^M\Vert_F^2
    \label{eqn:tuckeropt}
\end{equation}
where $\Vert\Vert_F$ is the {\em Frobenius norm} of the matrix.
The most popular approach to solve the above formulation is the {\em higher order orthogonal iteration} (HOOI) method, which uses the leading singular vectors of the matricized tensor along each mode to compute the factor matrices in an iterative fasion.
\subsubsection{CANDECOMP/PARAFAC (CP) Tensor Decomposition}
The CP decomposition~\cite{CEM:CEM582} method treats the tensor as a sum of finite number of rank one tensors (See Figure~\ref{fig:cp}).
\begin{defn}
  Let $\mathcal{T}$ be a tensor of size $N\times T\times M$. A rank-$R$ CP decomposition of $\mathcal{T}$ yields $R$ arrays for each mode, such that:
  \begin{equation}
    \mathcal{T} \approx \sum_{i=0}^{R}\lambda_i\times({\bf a}^N_{i}\circ {\bf a}^T_{i} \circ {\bf a}^M_{i})
    \label{eqn:cp}
  \end{equation} 
  where $\circ$ denotes the outer vector product. $R$ is referred to as the {\em rank} of the tensor.
  \label{defn:cp}
\end{defn}
For a given rank, $R$, computing the CP decomposition is typically done using an {\em Alternating Least Squares} (ALS) approach that minimizes the following objective function:
\begin{equation}
  \underset{\mathcal{T},{\bf A}^N,{\bf A}^T,{\bf A}^M}{\min} \Vert\mathcal{T} - \sum_{i=0}^{R}\lambda_i\times({\bf a}^N_{i}\circ {\bf a}^T_{i} \circ {\bf a}^M_{i})\Vert_F^2
  \label{eqn:cpopt}
\end{equation} 
where the vectors ${\bf a}^N_{i}, {\bf a}^T_{i}, {\bf a}^M_{i}$ are the $i^{th}$ columns of the matrices ${\bf A}^N,{\bf A}^T,{\bf A}^M$, respectively.
\subsection{Error Statistic for Tracking Performance}
Both Tucker and CP methods allow us to reconstruct a low rank approximation of the original tensor using~\eqref{eqn:tucker} and~\eqref{eqn:cp}, respectively. The low rank approximation captures the core behavior of the underlying system and ignores the noise, which allows for better performance tracking. We wish to extract a single statistic for each window, based on the low rank approximation. One obvious metric could be the recontruction error, defined for the tensor $\mathcal{T}_t$, as:
\begin{equation}
  \delta_t = \Vert\mathcal{T}_t - \mathcal{G}_t \times {\bf A}_t^M \times {\bf A}_t^N \times {\bf A}_t^T\Vert_F^2
  \label{eqn:tuckerreconstruct}
\end{equation}
where, $\mathcal{G}_t,{\bf A}_t^M,{\bf A}_t^N,{\bf A}_t^T$ are obtained via the Tucker decomposition of $\mathcal{T}_t$ for a specified rank, $(M',N',T')$. 

However, we use a slightly different procedure to compute the error statistic, as outlined in Algorithm~\ref{alg:tuckererror}.
\begin{algorithm}
  \begin{algorithmic}[1]
    \Function{TuckerError}{$\mathcal{T}_t, N',T', M'$}
    \State ${\bf P} \leftarrow Zeros(N,M)$
    \For{$m = 1$ to $M$}
    \State ${\bf P}[:,m] \leftarrow \mu_m$\Comment{Mean value for $m^{th}$ metric}
    \EndFor
    \State $\mathcal{Q} \leftarrow \mathcal{T}_t$
    \State $\mathcal{Q}[:,T,:] \leftarrow {\bf P}$\Comment{Replace the last time slice}
    \State $\mathcal{G},A_N,A_T,A_M \leftarrow Tucker(\mathcal{Q},N',T',M')$\Comment{See~\eqref{eqn:tuckeropt}}
    \State $\widehat{\mathcal{Q}} \leftarrow \mathcal{G}\times A_N \times A_T \times A_M$
    \State $\epsilon_t \leftarrow \Vert\widehat{\mathcal{Q}}[:,T,:] - \mathcal{T}_t[:,T,:]\Vert_F$
    \\
    \Return $\epsilon_t$
    \EndFunction

  \end{algorithmic}
  \caption{Tucker decomposition based error statistic}\label{alg:tuckererror}
\end{algorithm}

The error statistic, $\epsilon_t$, measures how well can the last time slice of the window be predicted using the first $T-1$ time slices. To achieve this, the observed values in the last time slice of the tensor is replaced with a ``dummy'' slice consisting of mean values obtained using first $T-1$ slices of the tensor. The Tucker decomposition followed by the reconstruction replaces the dummy entries with the predicted values.

The error statistic is then used to identify performance anomalies. The key hypothesis is that, given fixed rank and normal operating conditions, the capability of a tensor to make future predictions, measured as $\epsilon_t$, will be stable. Anomalies in the system performance will result in the error to become significantly larger than expected, and will be detected by inspecting the error statistic, $\epsilon_t, \forall t$.

The rank is a critical parameter in the analysis and is typically chosen by empirically studying the behavior of $\epsilon_t$ for different rank values. Clearly, if the rank is set to the size of the original tensor, i.e., $M' = M$, $N' = N$, and $T' = T$, $\epsilon_t$ will be zero. On the other, if the rank is minimal, i.e., $M' = N' = T' = 1$, then $\epsilon_t$ will be very high. However, we seek an optimal small rank, such that the error statistic $\epsilon_t$ is small but captures the behavior of the system during the given time window. This will allow for effective monitoring of the system behavior. 
\begin{figure*}[htbp]	
  \centering
  \subfloat[Raw $\epsilon_t$]{
    \includegraphics[width=0.5\textwidth]{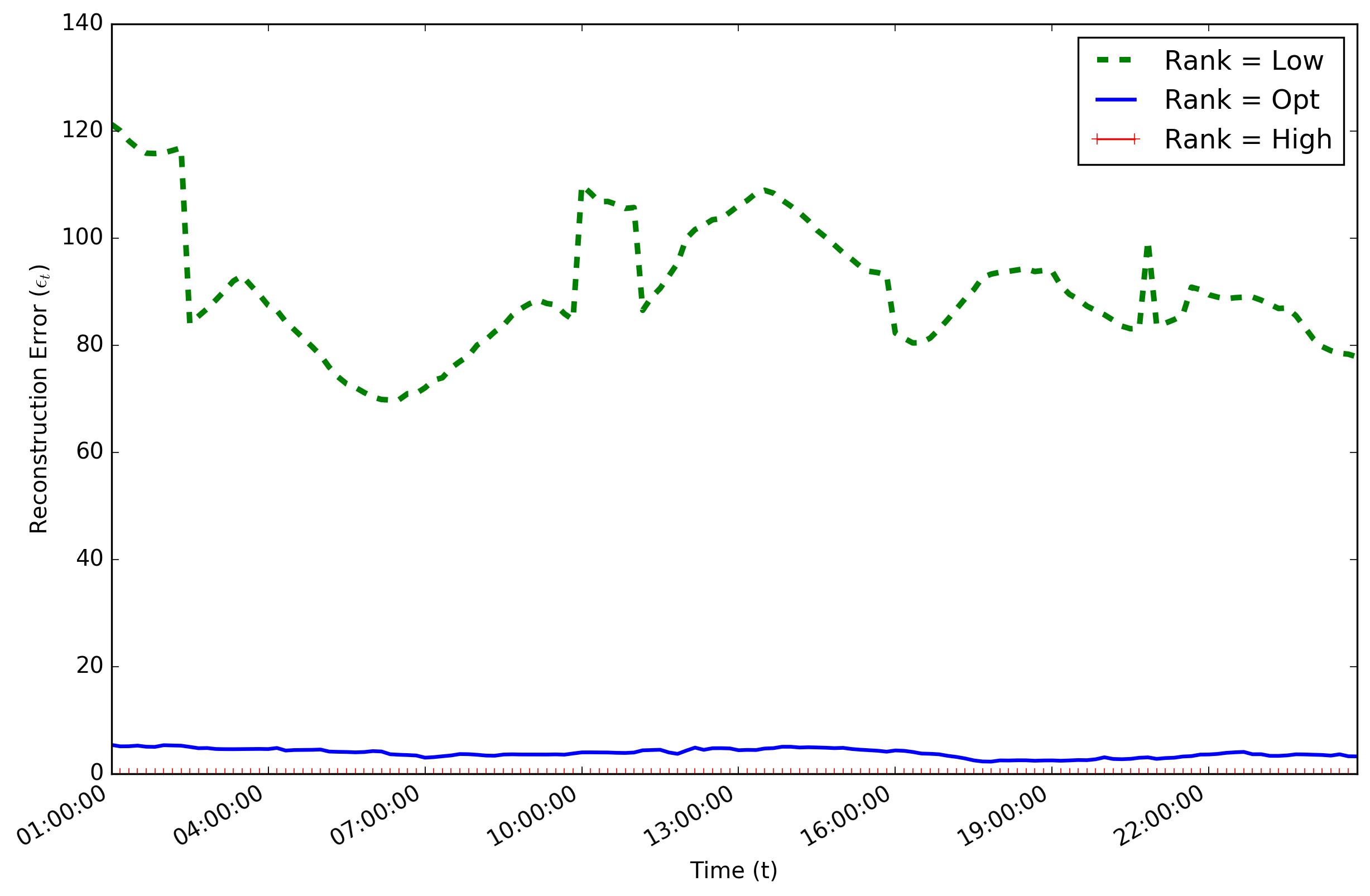}
    \label{fig:rankcomparisonraw}
  }
  \subfloat[Normalized $\epsilon_t$]{
    \includegraphics[width=0.5\textwidth]{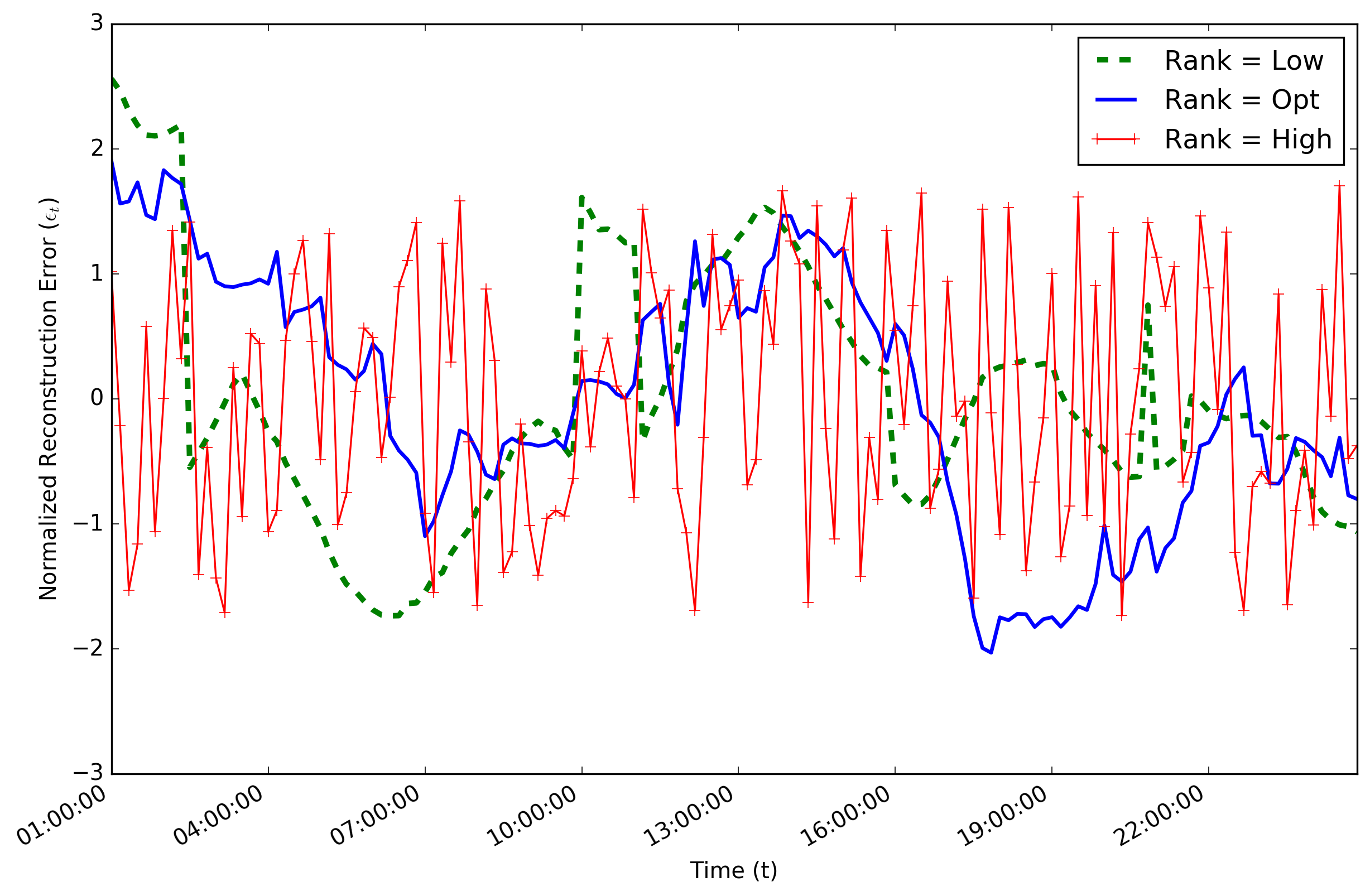}
    \label{fig:rankcomparisonnormalized}		
  }
  \caption{Comparing reconstruction error for different values of rank on a day-long Lonestar4 \tacc\ data (See \protect\ref{sec:setup}). Error is computed for a three hour long window (18 times slices). The rank values, i.e., $(N',T',M')$, for the three settings are: {\em low} - (1,1,1), {\em opt} - (300,18,30), and {\em high} - (1709,18,86), which is the full rank of the window.}
  \label{fig:rankcomparison}
\end{figure*}

For instance, Figure~\ref{fig:rankcomparison} compares the value of $\epsilon_t$ on Lonestar4 \tacc\ data for three different rank settings. When using full rank of the tensor for reconstruction, the error is nearly zero but does not capture any variance in the system behavior. Using the smallest possible rank exhibits higher variance along time, however, the error is high which means that the approximation poorly captures the information in the corresponding tensor, which will result in unreliable monitoring. An intermediate value of the rank provides a better tradeoff between the error and the variance over time. We will later show how to empirically determine the optimal rank, by studying the behavior of $\epsilon_t$.
\subsection{Using Job Information for a Refined Statistic}
For a given rank, the error statistic, $\epsilon_t$, is an indicator of the relative activity in the system. For a time window in which there are no running jobs, one would expect that the $\epsilon_t$ will be low, since all nodes would have similar resource usage during that time interval. On the other hand, if the job load is highly varied across nodes, the value of $\epsilon_t$ will be relatively higher. We use this knowledge to improve the predictive capability of the above approach, which in turn lowers the value of the error statistic.

%The core assumption here is that, under normal operations, the tensor at any given time can be reasonably approximated using Tucker (or CP) decomposition. In other words, $\epsilon_t$ is expected to be low. From the system perspective, this translates to the assumption that along each mode, the system behavior has correlations. For instance, along the metric mode, given that many metrics capture similar information, they could be represented as linear combinations of fewer latent metrics. However, for the mode representing the $N$ nodes, one would expect nodes to behave according to the jobs that are executed on that node. Which means that 
We use the job matrix available for the time window, denoted as ${\bf J}_t$ and defined in Section~\ref{sec:data}. In this representation, each node is described using a $J_t$ length binary vector, where $J_t$ is the number of jobs active in the system during that time window. The nodes are clustered using the data in ${\bf J}_t$ into $K$ clusters. While any clustering algorithm~\cite{Jain:1999} maybe employed for this task, we use the widely used {\em K-means clustering} algorithm, using {\em cosine distance} to compute distance between a pair of nodes. The cosine distance between nodes $i$ and $j$ is defined as:
\[
  cosine(i,j) = 1 - \frac{{\bf J}_t[i,:]\cdot{\bf J}_t[j,:]}{\Vert{\bf J}_t[i,:]\Vert_2\Vert{\bf J}_t[j,:]\Vert_2}
\]
The output of the clustering is a $N$ length vector, ${\bf c}_t$, such that each element of ${\bf c}_t$ indicates the cluster index (between 1 and $K$) to which the corresponding node is assigned by the clustering algorithm. The vector ${\bf c}_t$ is used to calculate the error statistic using the procedure outlined in Algorithm~\ref{alg:tuckererrorcluster}. 
The idea is to consider the data for nodes corresponding to each cluster separately. The sub-tensor corresponding to nodes in each cluster (computed in Line 5) is then passed to the {\tt TuckerError} algorithm (See Algorithm~\ref{alg:tuckererror}). The errors for each cluster are then combined to obtain the final statistic.
\begin{algorithm}
  \begin{algorithmic}[1]
    \Function{TuckerErrorCluster}{$\mathcal{T}_t,{\bf J}_t, N',T', M',K$}
    \State ${\bf c}_t \leftarrow KMeans({\bf J}_t,K)$
    \State $\epsilon_t \leftarrow 0$
    \For{$k=1$ to $K$}
    \State $\mathcal{T}_t^k \leftarrow \mathcal{T}[\{i:{\bf c}_t[i] = 1\},:,:]$
    \State $\epsilon^k_t \leftarrow TuckerError(\mathcal{T}_t^k, N',T', M')$
    \State $\epsilon_t \leftarrow \epsilon_t + (\epsilon^k_t)^2$
    \EndFor
    \State $\epsilon_t \leftarrow \sqrt{\epsilon_t}$
    \\
    \Return $\epsilon_t$
    \EndFunction

  \end{algorithmic}
  \caption{Tucker decomposition based error statistic (with clustering)}\label{alg:tuckererrorcluster}
\end{algorithm}

\begin{figure}[htbp]
  \centering
  \includegraphics[width=0.5\textwidth]{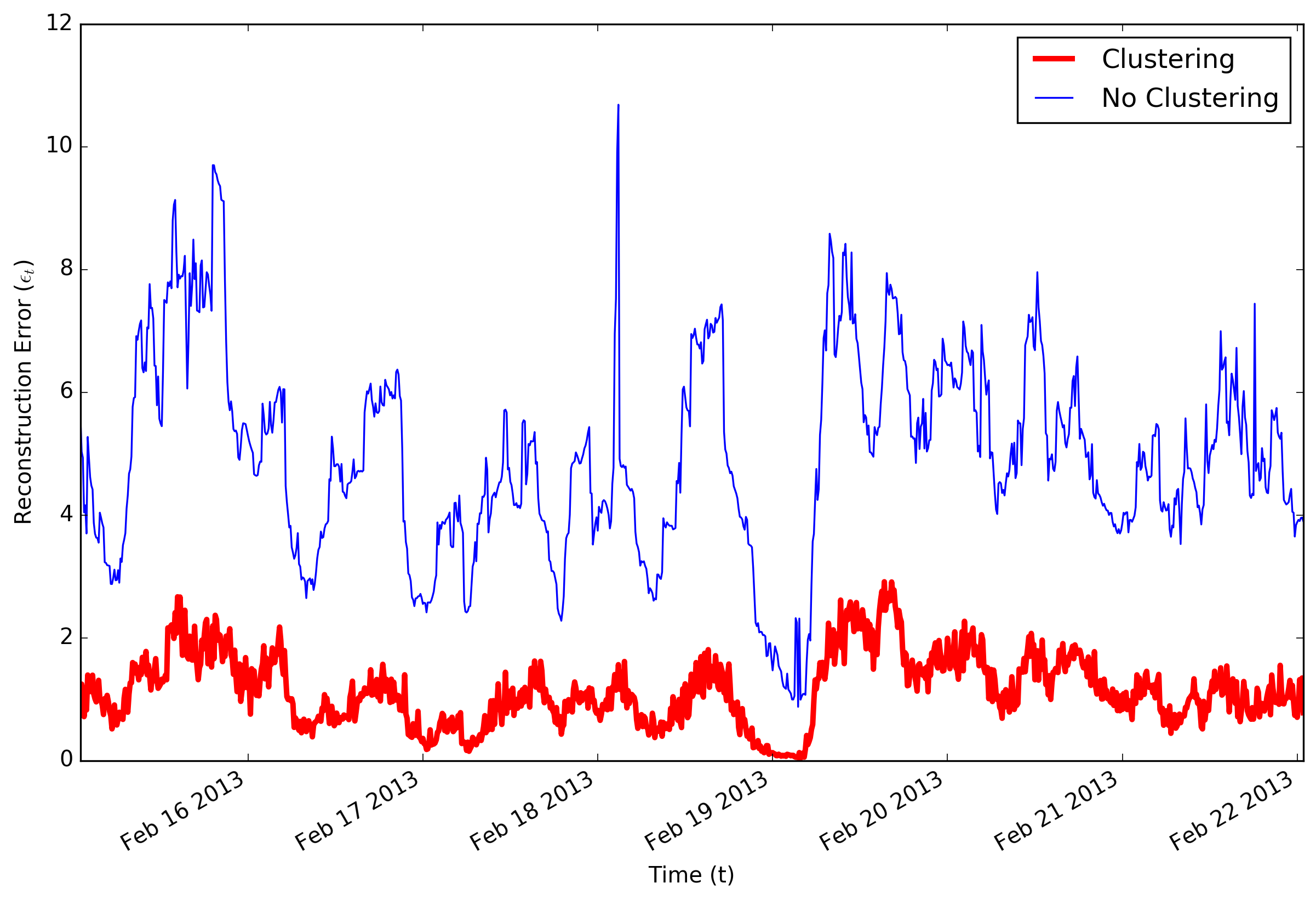}
  \caption{Comparing clustering vs.~no clustering performance}
  \label{fig:clustervsnoclustercomp}
\end{figure}

Intuitively, it is easy to see that the value of the error statistic using the clustering scheme will be lower than the error statistic computed, i.e.,:
\begin{align}
  TuckerErrorCluster(\mathcal{T}_t,{\bf J}_t,N',T',M',K) \le\nonumber\\TuckerErrorCluster(\mathcal{T}_t,{\bf J}_t,N',T',M',1)
  \label{eqn:errorcomparison}
\end{align}
for $K \ge 1$. 

Since each sub-tensor consists of resource usage data for nodes that have similar workloads, in terms of the running jobs, the low rank decomposition of the sub-tensor can better predict the sub-matrix within the last slice. In fact, one can arrive at an analytical proof to show that the above relation holds. The proof is not included in this paper. Figure~\ref{fig:clustervsnoclustercomp} shows the comparison for one week worth of the \tacc\ data.

The choice of the number of clusters, $K$, is an important parameter. Methods such as K-means expect $K$ to be specified. Other methods such as {\em mean-shift}~\cite{Comaniciu:2002} and {\em Affinity Propagation}~\cite{Frey:2007} can operate without specifying $K$. However, such methods have their own parameters which are equally challenging to specify. $K$ translates to the number of different possible workloads that exist for a given window, which could be inferred by studying the job specific information from the resource usage data. In this paper, we directly specify a fixed value for $K$. 

\section{Anomaly Detection}
\label{sec:anomalydetection}
The error statistic computed using Algorithm~\ref{alg:tuckererrorcluster} can be used as a single statistic to track the performance of the system, under the hypothesis that the value of $\epsilon_t$ can be reasonably modeled using a univariate time series model, under normal operations of the system. In case of an anomaly, the value of $\epsilon_t$ will deviate significantly from the underlying model. The model can be as simple as modeling $\epsilon_t$ as a Gaussian random variable with stationary mean and variance and using methods such as {\em Cumulative Sum} (CUSUM) or {\em Exponentially Weighted Moving Average} (EWMA), which are some of the classical methods for statistical process control~\cite{Montgomery:2001}. At the same time, the statistic also provides a visual output to track the system performance. For instance, the output in Figure~\ref{fig:clustervsnoclustercomp} shows an approximately half-day cyclic pattern in the behavior. Sudden spikes in this behavior can reveal performance anomalies, as will be discussed in Section~\ref{sec:results}.

\section{Data and Experimental Setup}
\label{sec:setup}
Experimental results are provided on data obtained from the (now inactive) Lonestar4 supercomputing cluster, situated at the Texas Advanced Computing Center. This cluster consists of $1888$ compute nodes from which $1766$ nodes are used for this study. The primary analysis is done on resource usage data for 2013 and the output is evaluated against system message logs obtained for the same time duration. 
\subsection{Resource Usage Data}
The resource usage data is collected using the \tacc\ system monitor~\cite{Evans2014}, which collects a range of usage related metrics from the overall system, including hardware performance counters, Lustre filesystem operation counts, and Infiniband device usage. Data is collected at each computational node, both periodically and at the beginning and end of a job.
\begin{table}[h]
\centering
\begin{tabular}{|p{1in}|p{2in}|}
\hline
{\bf Component} & {\bf Resource usage metrics}\\
\hline
\hline
CPU & user, nice, system, idle, iowait, irq, softirq\\
\hline
I/O&rd\_ios, rd\_merges, rd\_sectors, rd\_ticks, wr\_ios, wr\_merges, wr\_sectors, wr\_ticks\\
\hline
Lustre /scratch, /work & read\_bytes, write\_bytes, direct\_read, direct\_write, dirty\_pages, dirty\_pages, ioctl, open, close, mmap, seek, fsync, setattr, truncate, flock, getattr, statfs, alloc\_inode, setxattr, getxattr, listxattr, removexattr, inode\_permission, readdir, create, lookup, link, unlink, symlink, mkdir, rmdir, mknod, rename\\
\hline
Lustre network & tx\_msgs, rx\_msgs, rx\_msgs, tx\_bytes, rx\_bytes, rx\_bytes\\
\hline
Virtual Memory & pgpgin, pgpgout, pswpin, pswpout, pgalloc\_normal, pgfree, pgactivate, pgdeactivate, pgfault, pgmajfault, pgrefill\_normal, pgsteal\_normal, pgscan\_kswapd, pgscan\_direct, pginodesteal, slabs\_scanned, kswapd\_steal, kswapd\_inodesteal, pageoutrun, allocstall, pgrotated\\
\hline
\end{tabular}
\caption{List of resource usage metrics used for experiments.}
\label{tab:metrics}
\end{table}

Each file generated by the monitor is self-describing with a meta-data header and actual measurements grouped into records as shown in Table~\ref{tab:taccsample}.
\begin{table}[h]
  \centering
  {\footnotesize
    \begin{tabular}{|l|}
      \hline
      \texttt{tacc\_stats 1.0.3}\\
      \texttt{hostname c300-101.ls4.tacc.utexas.edu}\\
      \texttt{uname Linux x86\_64 2.6.18-194.32.1.el5\_TACC \#2 \ldots}\\
      \texttt{uptime 14043540}\\
      \texttt{block rd\_ios,E rd\_merges,E rd\_sectors,E,U=512B}\\ 
    \texttt{rd\_ ticks,E,U=ms ,\ldots}\\
      $\vdots$\\
      \texttt{1369285201 0}\\
      \texttt{block sr0 0 0 0 0 0 0 0 0 0 0 0}\\
      \texttt{block sda 49084176 677526 6343808434 188319052} $\ldots$\\
      \texttt{cpu 0 319172960 135 16110118 1068725347 235978} $\ldots$\\
      \hline
    \end{tabular}
    \caption{Meta-data (schema) description and sample data generated by \tacc.}
    \label{tab:taccsample}
  }
\end{table}
The data is available from January 1, 2013 to December 31, 2013 and contains 52,560 records per node, sampled once every 10 minutes. The records collected when a job starts or terminates were ignored for this analysis. In all, the \tacc\ monitor provides 216 performance metrics. We used 86 of these metrics which measure the resource usage for CPU, memory, and I/O systems. See Table~\ref{tab:metrics} for the list of usage metrics used in this paper. These are similar to the metrics chosen for similar analyses in previous works~\cite{Guan:2010}. Note that the {\tt cpu} related metrics were available for each processor, and a single average was used for each 10 minute block. For the metrics that record counters, we used the difference with the next block's counter value. The individual metrics were normalized to have zero mean and unit variance.
\subsection{System Log Messages}
While the proposed method operates on resource usage data, we use the system message log for the same time period to verify the identified events. This motivated from past studies that have shown correlations between events identified from resource usage data and event logs~\cite{Chuah:2013:LRU:2553409.2553428}. We use POSIX formatted log messages that include one message per line, consisting of the timestamp, host, protocol, and the event message. Few examples are shown below:\\
{\footnotesize
{\tt Feb 16 04:30:50 c318-116 kernel: [10716468.440192] Lustre: 14751:(file.c:3312:ll\_inode\_revalidate\_fini()) failure -2 inode 250689312 \\
Feb 12 01:28:42 c337-314 kernel: [10359976.556231] Lustre: 10883:(client.c:1476:ptlrpc\_expire\_one\_request() ) @@@ Request x1424903135350826 sent from work $\ldots$}}

For the year of 2013, a total of 642,988,509 log messages were available for the Lonestar4 system. 
\subsection{System Setup}
All experiments were performed on a Ubuntu Server with Intel(R) Xeon(R) CPU E5-2630 v3 @ 2.40GHz CPU, NVIDIA Corporation GK110GL [Quadro K5200] 8GB GPU and 16 GB memory. For the tensor decomposition we use the {\tt scikit-tensor} library~\cite{scikittensor} written in Python. For the Tucker based method in Algorithm~\ref{alg:tuckererror}, we avoid the high computational cost of reconstructing the tensor (Line 9), by utilizing the tensor manipulation routines in the {\tt Theano toolbox}~\cite{Theano:2016} which uses the GPU to accelerate the computation. The matrix and vector operations were done using the Python {\tt numpy} library.

\section{Results}
\label{sec:results}
In this section we show the effectiveness of using the proposed error statistic in tracking the performance of an HPC system. Additionally we show visual inspection can reveal anomalies which are then verified to be meaningful by comparing with the system log messages. The data is from the Lonestar4 system, as described in Section~\ref{sec:setup}.
\subsection{Optimal Parameters}
\begin{figure}
  \centering
  \includegraphics[width=0.5\textwidth]{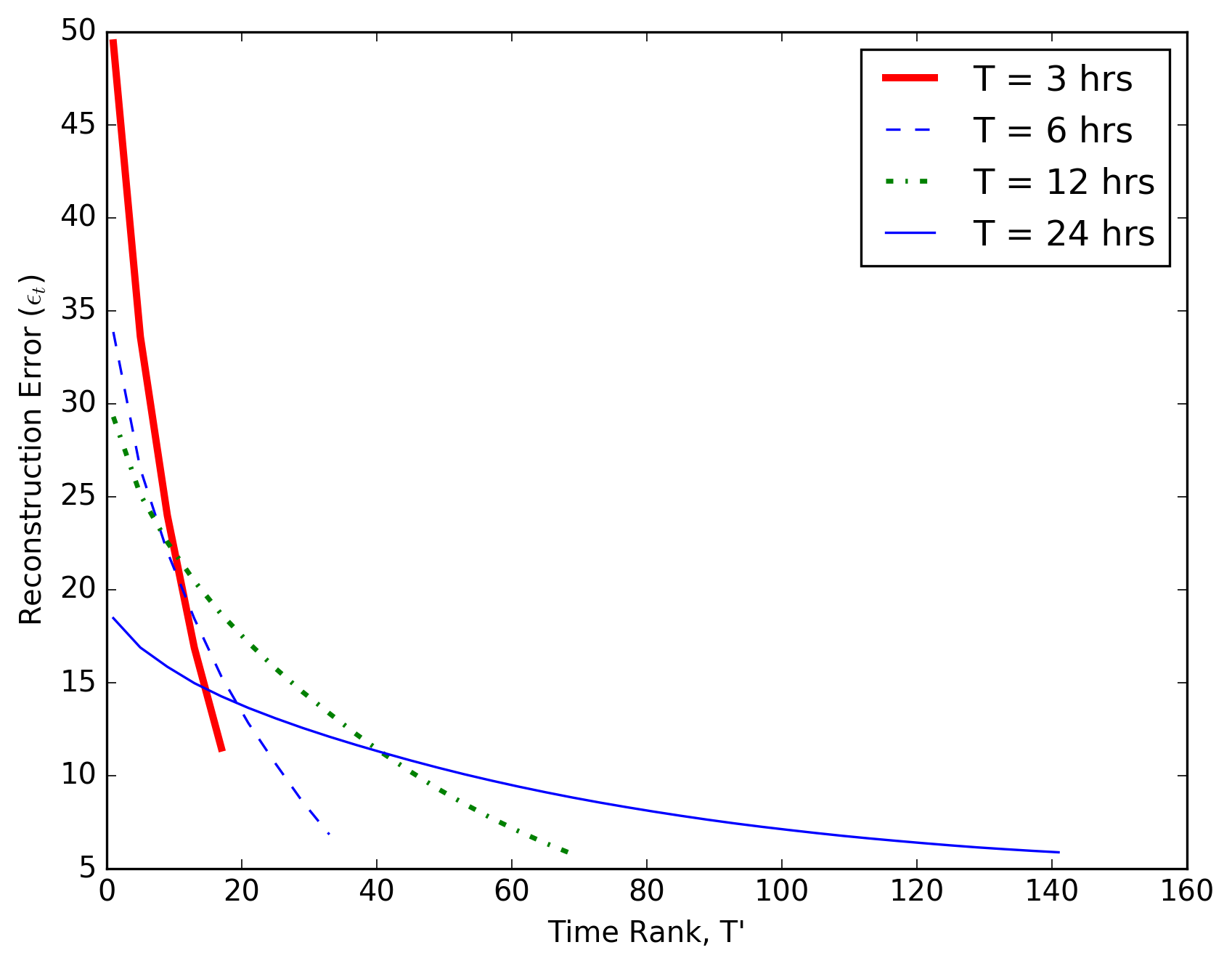}
  \caption{Relationship between error statistic, $\epsilon_t$, and time rank, $T'$, for different window sizes, $T$. For each window size, the rank is varied from 1 to $T$. The metric rank and node rank are kept fixed at 86 (full) and 300, respectively.}
  \label{fig:comparewindowsizes}
\end{figure}

There are five parameters required to estimate the error statistic in Algorithm~\ref{alg:tuckererrorcluster}. These are the lower rank for tensor decomposition, i.e., $(N',T',M')$, the length of the window, $T$, and the number of clusters, $K$. As discussed earlier, we fix the parameter $K$ in this paper, though one can infer optimal $K$ for a given time window based on the expected number of job usage patterns in the system. In these experiments we set $K = 5$.
\paragraph{Setting Optimal Window Size}
To understand the impact of window size, $T$, on the analysis, we conducted the following experiment. For a given window size and a fixed node and metric rank, i.e., $N'$ and $M'$, we varied the time rank, $T'$ from 1 to $T$ (full rank) and observed how the error statistic ($\epsilon_t$) varies with the time rank for a randomly chosen window. This allows us to understand the dependence among nodes and metrics along the time mode. The results are shown in Figure~\ref{fig:comparewindowsizes}. Ideally, in the presence of correlations along the time dimension, one would expect the error curve to decline sharply towards a low value and then stabilize.

\begin{figure}
  \centering
  \includegraphics[width=0.5\textwidth]{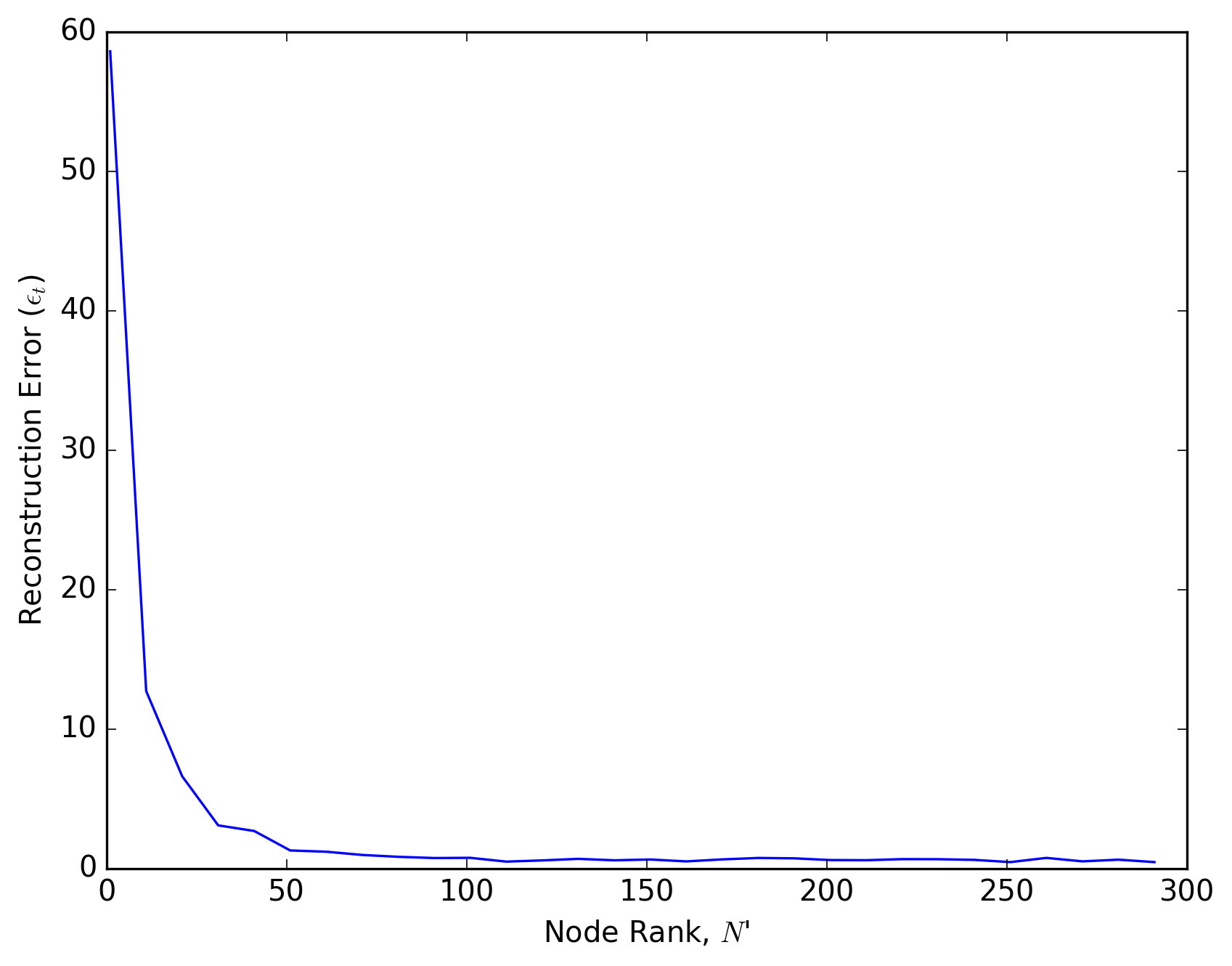}
  \caption{Relationship between error statistic, $\epsilon_t$, and node rank, $N'$. The metric rank and time rank are kept fixed at 30 and 18 (full), respectively.}
  \label{fig:comparenoderanks}
\end{figure}

However, as observed in Figure~\ref{fig:comparewindowsizes}, the value of the error statistic decreases almost linearly until rank is increased to 18 (corresponding to 18 10-minute slices or 3 hours). This means that within a 3 hour window, the system does not exhibit significant correlation along the time mode. Moreover, even if the window size is greater than 3 hours, one needs to consider a rank of more than 18 to get low error. For this reason, we choose the window size $T$ to be equivalent to 3 hours of observations.
\paragraph{Setting Optimal Rank}
For the chosen window size of 3 hours, the error statistic decreases linearly with increasing time rank (See Figure~\ref{fig:comparewindowsizes}). Thus we set $T'$ to be the full rank along the time mode, i.e., $T' = 18$. 
\begin{figure}
  \centering
  \includegraphics[width=0.5\textwidth]{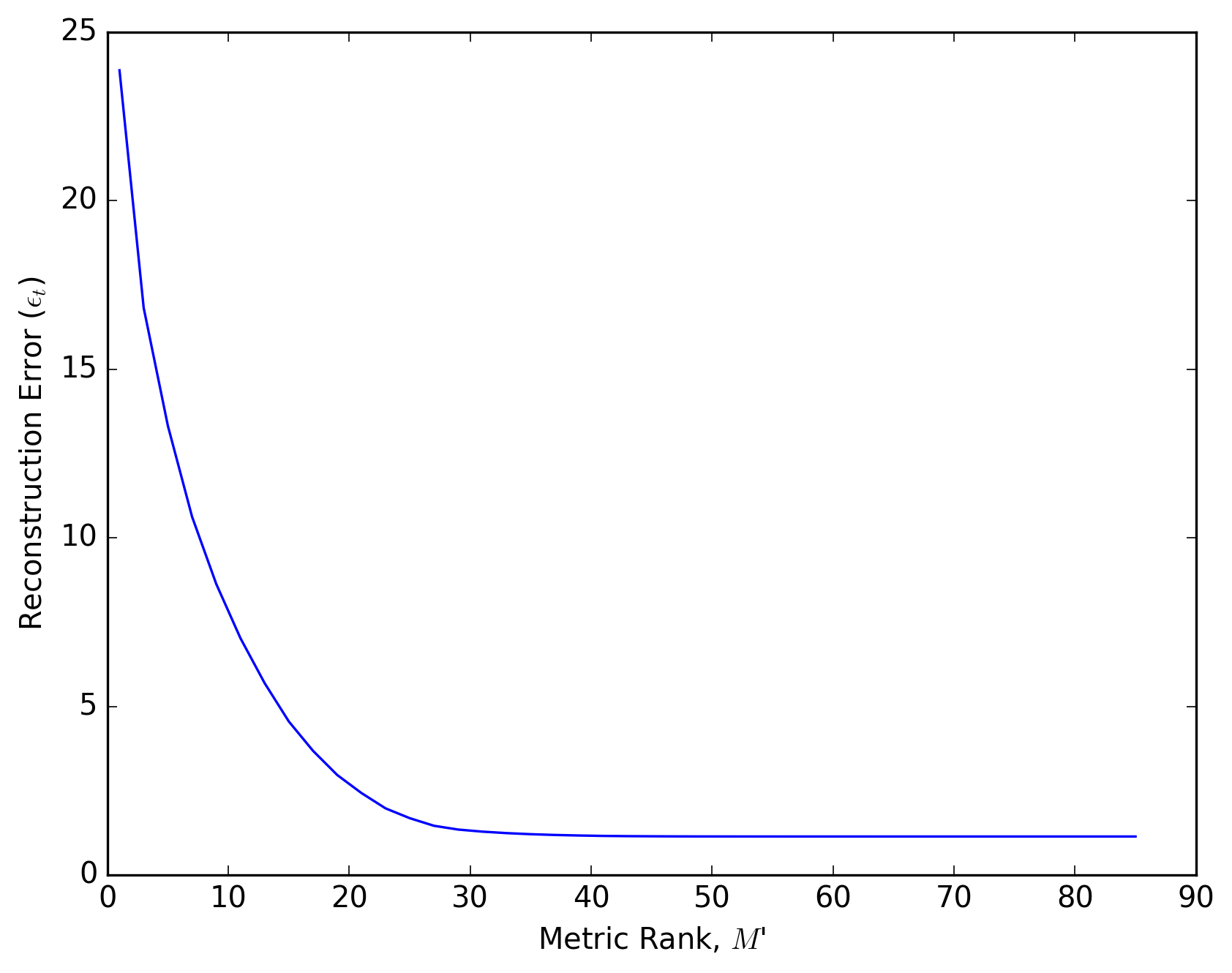}
  \caption{Relationship between error statistic, $\epsilon_t$, and metric rank, $M'$. The node rank and time rank are kept fixed at 300 and 18 (full), respectively.}
  \label{fig:comparemetricranks}
\end{figure}
Figure~\ref{fig:comparemetricranks} shows that a metric rank of 30 is enough to capture most of the information in the data. In the experiments, we set $M'=30$. Similarly, the node rank, $N'$, is set to 50, based on the behavior of the error statistic with varying $N'$, as shown in Figure~\ref{fig:comparenoderanks}. 
\subsection{Using Error Statistic for Identifying Performance Anomalies}
\begin{figure*}[ht]
  \centering
  \includegraphics[width=0.95\textwidth]{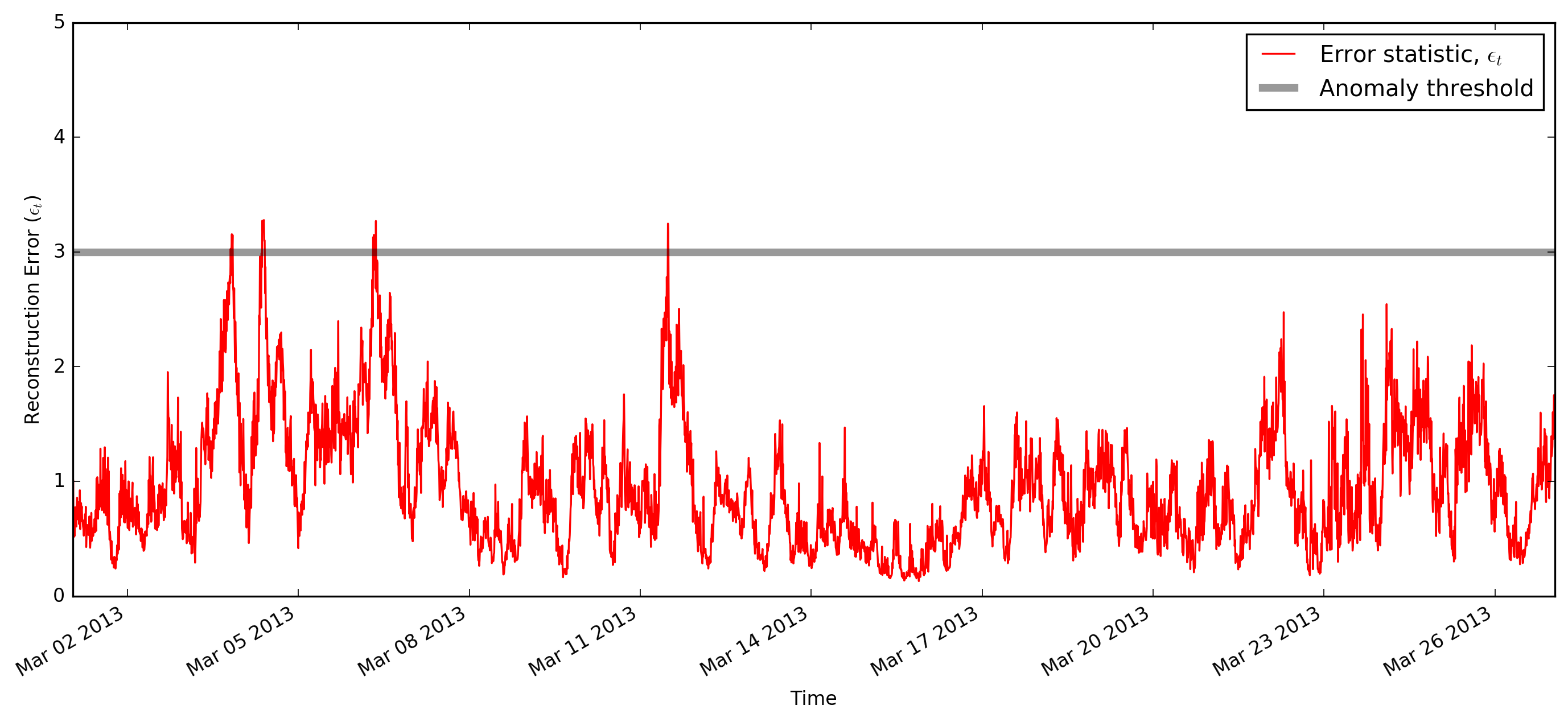}
  \caption{Using error statistic, $\epsilon_t$ to track performance of the Lonestar4 cluster for March 2013. Also shown is a user-supplied threshold to identify performance anomalies. The four anomalies that lie above the threshold are verified using the system log data.}
  \label{fig:marchresults}
\end{figure*}

Using the optimal parameters discussed above, we analyzed several weeks of the Lonestar4 data to understand the system performance. Figure~\ref{fig:marchresults} show the error statistic for four weeks in March, 2013. We choose a threshold of 3 to identify performance anomalies from this data, which gives us four anomalous events during the month. In general, one can study the statistical properties of $\epsilon_t$ to automatically determine an appropriate threshold for identifying anomalies.

Table~\ref{tab:errors} lists the critical errors from the system log data for the time windows corresponding to the four anomalous events identified in Figure~\ref{fig:marchresults}. We observe that the system reported several critical error messages during each of the four identified events, confirming that the proposed error statistic is able to identify truly anomalous events. For comparison, we also analyze the error messages for time windows with low value of $\epsilon_t$, such as the last row in Table~\ref{tab:errors}. While some error messages occur in those windows as well, e.g., segfaults, errors such as unreachable nodes only occur for the anomalous time windows.
\begin{table*}[ht]
  \centering
  \begin{tabular}{|c|c|p{4in}|}
    \hline
    Time & Type & Errors\\
    \hline
    Mar 4, 23:50 - Mar 5, 00:40 & Anomalous & Lustre inode failure (kernel), CRITICAL host unreachable, segfault (kernel)\\
    Mar 5, 14:20 - Mar 5, 15:20 & Anomalous & Lustre inode failure (kernel), segfault (kernel), lustre communication error (kernel)\\
    Mar 7, 15:00 - Mar 7, 16:00 & Anomalous & CRITICAL host unreachable, Lustre write error (kernel), Lustre inode failure (kernel)\\
    Mar 13, 13:10 - Mar 13, 13:20& Anomalous & CRITICAL host unreachable, Lustre inode failure (kernel), Lustre write error\\
    \hline
    Mar 17, 16:00 - Mar 17, 17:00& Normal & Lustre inode failure (kernel), segfault (kernel)\\
\hline
  \end{tabular}
  \caption{Analysis of system log errors for time windows corresponding to anomalous resource usage event. Also shown is errors for one normal time window for comparison.}
  \label{tab:errors}
\end{table*}

\section{Related Works}
\label{sec:related}
We briefly discuss related research in two areas that are relevant to the proposed methodology. First is the area of system monitoring using log data and second is the area of tensor based monitoring of non-HPC systems.
\paragraph*{HPC system performance monitoring methods}
Solutions for detecting, diagnosing, and predicting faults and failures in large high performance computing installations have typically relied on message logs~\cite{Xu:2008,Oliner:2008,Fu:2009,Reidemeister:2009,Pelaez:2014} or resource usage data~\cite{Guan:2010,Bronevetsky:2012}, or both~\cite{Chuah:2013:LRU:2553409.2553428,Gurumdimma:2016,Chuah:2016}. Since this paper focuses on the detection task, we present a brief overview of related methods that deal with detecting faults. Methods operating on message logs typically aggregate message logs by an entity of interest, e.g., a computational node/block~\cite{Oliner:2008,Xu:2008} or a job~\cite{Fu:2009}, and identify anomalous entities using different data representations, such as message arrival statistics~\cite{Oliner:2008}, vector representation derived from the message content~\cite{Xu:2008} or a state machine that models the dynamic behavior of the entity~\cite{Fu:2009}. However, message logs are typically noisy and often incomplete, which has led to methods that analyze alternate data sources such as resource usage metrics or performance counter data~\cite{Guan:2010,Bronevetsky:2012}. However, these solutions perform a node-specific or job-specific analysis of resource usage to identify anomalous nodes~\cite{Guan:2010} or jobs~\cite{Bronevetsky:2012}. Recently, solutions that combine resource usage data and message logs to improve fault detection have been proposed~\cite{Gurumdimma:2016,Chuah:2016}. The Crude system uses resource usage data to improve the performance of a PCA driven anomaly detection method~\cite{Xu:2008} that operates at node and job level. However, none of the existing methods model the temporal dimension to better identify faults.. 
\paragraph*{Tensor decomposition based monitoring in other domains}
In this paper we model the resource usage data as a 3-way tensor. While, a tensor based approach has not been previously employed for system monitoring, related approaches have been explored in other contexts, including fundamental work in using tensor decomposition techniques for mining multi-aspect data~\cite{4781131}. In particular, methods that use the reconstruction error between the raw data tensor and the tensor recovered from the low-rank representation have been applied for identifying anomalies in hyper-spectral imagery~\cite{Zhang:2016} and surveillance from spatio-temporal sensor measurements~\cite{DBLP:journals/corr/Fanaee-TG14a}. The latter work~\cite{DBLP:journals/corr/Fanaee-TG14a} deals with time as one of the tensor dimensions, similar to our work. However, the approach tracks the changes in the subspace from one temporal ``slice'' (as a 2-way matrix) to the next. In our proposed work, we use a fixed size historical tensor to predict the next one or more temporal slices. 
%Tensor decomposition based methods have been explored in other contexts. Tensors are multi-array systems. As tensors can be $1-D, 2-D, . . . , and, N-D$ order, $1-D$ tensor can be called as array, $2-D$ tensor can be called as matrix. Multivariate spatio-temporal tensor analysis can be analyzed with low rank tensor learning by using fast-greedy tensor learning related to the specified rank.~\cite{NIPS2014_5429}, 
%missing data can be revealed, imputed by using low dimensional subspace and a subspace estimator using exponentially weighted least squares.~\cite{7072498}, Decomposing 3-way tensors, Tucker method  generates three factor matrices and a core tensor for the specified ranks for each of the orders. Anomaly detection is identified at the tensor by calculating the error between the original tensor and tensor, slice or fiber reconstruction followed by the tensor decomposition. ~\cite{4781131}. Deviations and changes in multi array syndromic surveillance data can be tracked without considering the use of top-down or bottom-up search scheme and eigenspace techniques would enhance the anomalous activities without increasing the false alarm rates.~\cite{DBLP:journals/corr/Fanaee-TG14a}  

\section{Conclusions and Future Directions}
\label{sec:conclusions}
The proposed approach performs system level monitoring and provides a coarse assessment of the global behavior of an HPC system. The results on Lonestar4 data have shown that the method effectively captures the relationships across usage metrics, computational nodes, and time to extract a single temporal signal which can be used by system administrators to monitor the system health.

In future, we plan to integrate the information from system messages into the analysis, rather than using that for post-validation. Other researchers have reported promising results by analyzing the two sources of information together~\cite{Gurumdimma:2016,Chuah:2016}. While the proposed error statistic allows for a visual tracking of the system performance, we plan to build a statistical anomaly detection model that leverages the expected behavior of the statistic to identify appropriate thresholds for flagging an anomalous event. Moreover, while the statistic by itself allows for only identifying anomalies, one can inspect the low rank decomposition to explain the anomalies, and will be explored in near future.% In particular, we plan to use annotated messages that provide job and node information for each error message, which can be used to construct a parallel tensor to the resource usage tensor.
%Tensor decomposition methods provide relative information on the system behaviour. On complex structure of high performance computing systems evaluating the system behavior is possible using the system counter data and the system log message counts to some extend. Whenever HPC systems use more resources or they behave strangely there is some anomolous behaviour happening. Anomolous behaviors cause systems performing poorly and that can be indicated from the system log message increase or system component counter data values. Log message count and the system counter data correlate to some extent. 

\section*{Acknowledgements}
{\small
Authors wish to acknowledge support provided by the Center for Computational Research at the University at Buffalo. Additionally, the authors thank researchers at Texas Advanced Computing Center for access to the Lonestar4 data and related support.
}
\bibliographystyle{abbrv}
\bibliography{references} 
\end{document}